\documentstyle[12pt]{article}
\raggedright  \parindent=20pt

\title{ Measuring velocity ratios \\ with correlation functions }
\author{ David Seibert\thanks{On leave until 12 October 1993 at:
Theory Division, CERN, CH--1211 Geneva 23, Switzerland; Internet:
seibert@surya11.cern.ch.} \\
{\it Department of Physics } \\ {\it Kent State University, Kent, OH
44242 USA } \\ Kevin Haglin\thanks{Internet:
haglin@hep.physics.mcgill.ca.}~~and Charles Gale\thanks{Internet:
gale@hep.physics.mcgill.ca.} \\ {\it Department of Physics } \\
{\it McGill University, Montr\'eal, P.Q., H3A 2T8, Canada } }

\begin{document}

\maketitle

\begin{abstract}
We show how to determine the ratio of the transverse velocity of a
source to the velocity of emitted particles, using split--bin
correlation functions.  The technique is to measure $S_2$ and
$S_2^{\phi}$, subtract the contributions from the single--particle
distribution, and take the ratio as the bin size goes to zero.  We
demonstrate the technique for two cases: each source decays into two
particles, and each source emits a large number of particles.
\end{abstract}

\leftline{PACS Indices: 25.70.Np, 12.38.Mh, 12.40.Ee, 24.60.Ky.}

\vfill \eject

\section{ Introduction }

There has been considerable study of correlation functions in high
energy and nuclear physics.  However, with the exception of source size
measurements using identical particle (Hanbury--Brown/Twiss)
interferometry [\ref{rhbt}], the connection between measurements of
correlation functions and the physics of particle collisions is tenuous.
For example, it is possible to measure the size of correlation sources
(the number of pions produced) in high energy collisions [\ref{rsizes}],
but no other source characteristics are currently identifiable.  This is
unfortunate, as the source size is anomalously large in
ultra--relativistic nuclear collisions [\ref{rsizes}, \ref{ranom}] (the
sources decay into at least fifteen pions each [\ref{rdrops}]),
and the nature of these sources has not yet been determined.

In this article, we propose to measure the velocities of correlation
sources, the objects that make up the intermediate structure of
collisions, by using the rapidity, $y$, and azimuthal angle, $\phi$,
correlation functions [\ref{rsfms}]
\begin{equation}
S_2(\delta y; \Delta Y) ~=~ \frac {\displaystyle \left\langle
\rho^{(2)} \right\rangle_{\delta y}} {\displaystyle \left\langle
\rho^{(2)} \right\rangle_{\Delta Y}} ~, \label{es2}
\end{equation}
and
\begin{equation}
S_2^{\phi}(\delta y; \Delta Y) ~=~ \frac {\displaystyle
\left\langle \rho^{(2)} \right\rangle_{\delta y}^{\phi}} {\displaystyle
\left\langle \rho^{(2)} \right\rangle_{\Delta Y}^{\phi}}~, \label{es2phi}
\end{equation}
where measurements are made over a window from $Y_0$ to $Y_0+\Delta Y$.
Here
\begin{equation}
\left\langle \rho^{(2)} \right\rangle_x ~=~
\sum_{i=1}^{\Delta Y/x} \frac {\displaystyle
\int_{Y_0+(i-1)x}^{Y_0+(i-1/2)x} \!\!\! dy_1
\int_{Y_0+(i-1/2)x}^{Y_0+ix} \!\!\! dy_2 \, \rho^{(2)}}
{\displaystyle x \Delta Y/4} \, ,
\end{equation}
and
\begin{equation}
\left\langle \rho^{(2)} \right\rangle_x^{\phi} ~=~
\sum_{i=1}^{\Delta Y/x} \frac {\displaystyle
\int_{Y_0+(i-1)x}^{Y_0+ix} \!\!\! dy_1
\int_0^{\pi} \!\!\! d\phi_1^{\rm lab}
\int_{Y_0+(i-1)x}^{Y_0+ix} \!\!\! dy_2
\int_{\pi}^{2\pi} \!\!\! d\phi_2^{\rm lab} \, \rho^{(2)}}
{\displaystyle x \Delta Y/4} \, ,
\end{equation}
where $\Delta Y$ is an integer multiple of $x$.
Split--bin correlation functions are useful for measuring correlations
with maximum statistics (without re--use of data).  $S_2$ is closely
related to the standard two--particle correlation function:
\begin{equation}
S_2 (\delta y; \Delta Y) ~\simeq~ \frac {\rho^{(2)} (0, \delta y/2)}
{\rho^{(2)} (0, \Delta Y/2)}~.
\end{equation}
For large $\Delta Y$, the correlation functions are
almost independent of $\Delta Y$, but in practice it is impossible to
make correlation measurements in particle collisions without using a
large bin for reference, as the amount of produced matter fluctuates
from event to event.

The proposed technique is to measure the ratio
\begin{equation}
R (\Delta Y) ~=~ \frac {S_2^{\phi} (0; \Delta Y) \, - \, \Sigma_2^{\phi}
(\Delta Y)} {S_2 (0; \Delta Y) \, - \, \Sigma_2 (\Delta Y)}~.
\label{eR0} \end{equation}
Here
\begin{equation}
\Sigma_2 (\Delta Y) ~=~ \frac {\displaystyle \Delta Y \,
\int_{Y_0}^{Y_0+\Delta Y} \!\!\! dy \, \rho^2 (y)}
{\displaystyle 4 \, \int_{Y_0}^{Y_0+\Delta Y/2} \!\!\! dy_1
\int_{Y_0+\Delta Y/2}^{Y_0+\Delta Y} \!\!\! dy_2 \, \rho (y_1) \,
\rho (y_2)} \label{esig2}
\end{equation}
and
\begin{equation}
\Sigma_2^{\phi} (\Delta Y) ~=~ \frac {\displaystyle \Delta Y \,
\int_{Y_0}^{Y_0+\Delta Y} \!\!\! dy \int_0^{\pi} \!\!\! d\phi_1^{\rm lab}
\int_{\pi}^{2\pi} \!\!\! d\phi_2^{\rm lab} \, \rho (y, \phi_1^{\rm lab})
\, \rho (y, \phi_2^{\rm lab})}
{\displaystyle \int_{Y_0}^{Y_0+\Delta Y} \!\!\! dy_1
\int_0^{\pi} \!\!\! d\phi_1^{\rm lab}
\int_{Y_0}^{Y_0+\Delta Y} \!\!\! dy_2
\int_{\pi}^{2\pi} \!\!\! d\phi_2^{\rm lab} \,
\rho (y_1, \phi_1^{\rm lab}) \,
\rho (y_2, \phi_2^{\rm lab})}~, \label{esig2phi}
\end{equation}
are the values of $S_2$ and $S_2^{\phi}$ from the single--particle
distributions, $\rho (y)$ and $\rho (y, \phi^{\rm lab})$.  In the case
of a flat distribution [$\rho (y, \phi^{\rm lab}) = {\rm const.}$],
$\Sigma_2 = \Sigma_2^{\phi} = 1$.
It is also possible to measure emission velocities using standard
two--particle interferometry techniques; however, these measurements
are somewhat more difficult than the technique proposed here, as they
require particle identification.  The technique proposed here has
the advantage of being useable even in the absence of particle
identification.

We assume isotropic emission and show that, in the non--relativistic
limit, $R$ depends only on the velocity ratio, $v_s/v_{\pi}$, where
$v_s$ is the source transverse velocity, and $v_{\pi}$ is the velocity
of emitted particles in the source rest frame.  In section~\ref{stpd},
we calculate $R$ for the case of a resonance that decays isotropically
into two particles.  In section~\ref{sts}, we calculate $R$ for
isotropic decay of a thermal source, finding that the shape is not very
different from the first case.  Finally, we summarize our results and
discuss applications and future work in section~\ref{sc}.

\section{ Two--particle decay } \label{stpd}

Suppose that we have an event with a single resonance that decays into
two particles, and a flat background of $N-2$ particles in a window
of rapidity $\Delta Y$.  The background contribution to the
two--particle density is then constant:
\begin{equation}
\rho^{(2)}_b (y_1, \phi_1^{\rm lab}, y_2, \phi_2^{\rm lab}) ~=~
(N-2) \, (N-3) \, / \, (4 \, \pi^2 \, \Delta Y^2)~. \label{er2b}
\end{equation}
Adding the contribution from combining one background particle with one
particle from the resonance, we obtain the total uncorrelated signal,
\begin{equation}
\rho^{(2)}_u (y_1, \phi_1^{\rm lab}, y_2, \phi_2^{\rm lab}) ~=~
\left[N \, (N-1) \, - \, 2\right] \, / \, (4 \, \pi^2 \, \Delta Y^2)~.
\label{er2u} \end{equation}

Given the resonance azimuthal angle, $\phi_r$, and the resonance
rapidity, $y_r$, we calculate the laboratory coordinates, $y$ and
$\phi^{\rm lab}$, as functions of the center-of-momentum coordinates,
$\theta$ and $\phi$.
\begin{eqnarray}
y ~=~ \tanh^{-1} \left( \frac {(1-v_s^2)^{1/2} \, v_{\pi} \,
\cos \theta} {1 ~+~ v_s \, v_{\pi} \, \sin \theta \, \sin \phi}
\right) ~+~ y_r~, \label{eylab} \\
\phi^{\rm lab} ~=~ \tan^{-1} \left( \frac {v_s ~+~
v_{\pi} \, \sin \theta \, \sin \phi} {(1-v_s^2)^{1/2} \, v_{\pi} \, \sin
\theta \, \cos \phi} \right) ~+~ \pi \, H (-\cos \phi) ~+~ \phi_r~,
\label{eplab} \end{eqnarray}
where
\begin{equation}
H(x) ~=~ \left\{ \begin{array}{ll}
                    0\quad & x \leq 0, \\
                    1\quad & x > 0.
                 \end{array} \right.
\end{equation}
[We use standard high--energy units, with $c=1$.  In deriving
eq.~(\ref{eplab}), we assume that $-\pi/2 \, \leq \, \tan^{-1} x \, < \,
\pi/2$.]  We assume that $y_r$ and $\phi_r$ are distributed randomly,
with flat distributions.  The resonance contribution to $\rho^{(2)}$ is
simplest in the resonance rest frame:
\begin{equation}
\rho^{(2)}_{\rm r} (\theta_1, \phi_1, \theta_2, \phi_2) ~=~ \frac
{\sin \theta_1 \, \sin \theta_2 \, \delta \left( \cos \theta_1 \, + \,
\cos \theta_2 \right) \, \delta \left( \left| \phi_1-\phi_2 \right| \,
- \, \pi \right)} {2 \, \pi},
\end{equation}
where $\delta$ is the Dirac $\delta$--function.

The resonance contribution to $S_2(\delta y; \Delta Y)$ is
\begin{equation}
S_{2, \rm r} (\delta y; \Delta Y) = \frac
{\displaystyle \Delta Y \int_0^{\Delta Y} \!\!\! dy_r
\int_0^{\delta y/2} \!\!\! dy_1 \int_{\delta y/2}^{\delta y} \!\!\! dy_2
\int_0^{2\pi} \!\!\!\!\! d\phi_1 \int_0^{2\pi} \!\!\!\!\! d\phi_2
\rho^{(2)}_{\rm r} (y_1, \phi_1, y_2, \phi_2)} {\displaystyle \delta y^2
\, \pi^2 \, \Delta Y^2 \, \rho^{(2)}_u (y_1, y_2)}, \label{es2a}
\end{equation}
where
\begin{equation}
\rho_{\rm r}^{(2)} (y_1, \phi_1, y_2, \phi_2) ~=~
\left| \frac {\partial \theta_1} {\partial y_1} \right| \,
\left| \frac {\partial \theta_2} {\partial y_2} \right| \,
\rho_{\rm r}^{(2)} (\theta_1, \phi_1, \theta_2, \phi_2)~. \label{erho2g}
\end{equation}
As we assume flat distributions, the contribution from all bins is the
same, so we use only the bin from $y=0$ to $\delta y$, and take
$Y_0=0$  The contribution at $\delta y = 0$ is
\begin{eqnarray}
S_{2, \rm r} (0; \Delta Y) =
\lim_{\delta y \rightarrow 0} \frac {\displaystyle 4 \Delta Y
\left[ \int_{\delta y/4}^{\delta y/2} dy_r \int_0^{2y_r-\delta y/2}
\!\!\! dy_1 + \int_{\delta y/2}^{3\delta y/4} dy_r
\int_{2y_r-\delta y}^{\delta y/2} \!\!\! dy_1 \right]}
{\displaystyle \delta y^2 \, \left[N \, (N-1) \, - \, 2\right] \,
\left[ (1-v_s^2)^{1/2} \, v_{\pi} \right]}, \label{es2d} \\
=~ \frac {\displaystyle \Delta Y}
{\displaystyle 2 \, \left[N \, (N-1) \, - \, 2\right] \,
(1-v_s^2)^{1/2} \, v_{\pi}}. \label{es2e}
\end{eqnarray}

Evaluating $S_{2, \rm r}^{\phi}$ in the same manner, we obtain
\begin{equation}
S_{2, \rm r}^{\phi} (0; \Delta Y) = \frac {\displaystyle \Delta Y
\int_0^{2\pi} \!\!\!\!\! d\phi_r
\int_{\phi_1^{\rm lab}=0}^{\pi} \!\!\!\!\!\! d\phi_1
\int_{\phi_2^{\rm lab}=\pi}^{2\pi} \!\!\!\!\!\! d\phi_2
\delta \left( \left| \phi_1-\phi_2 \right| - \pi \right)}
{2 \, \pi^2 \, [N(N-1) - 2] \, (1-v_s^2)^{1/2} \, v_{\pi}}~.
\label{eS2pd} \end{equation}
Note that $\phi^{\rm lab}$ is now independent of $\theta$, as in the
limit $\delta y \rightarrow 0$ the only contribution is from
$\cos \theta=0$.  We then perform the integral over $\phi_r$.
\begin{equation}
S_{2, \rm r}^{\phi} (0; \Delta Y) = \frac {\displaystyle \Delta Y
\int_0^{2\pi} d\phi \,
g \left( \left| \phi^{\rm lab} (\phi) \, - \, \phi^{\rm lab} (\phi+\pi)
\right| \right)}
{2 \pi^2 \, [N(N-1) - 2] \, (1-v_s^2)^{1/2} \, v_{\pi}}~,
\label{eS2pe} \end{equation}
where
\begin{equation}
g(x) ~=~ \left\{ \begin{array}{ll}
                 x\quad & 0 \, \leq \, x \, < \, \pi~, \\
                 2\pi \, - \, x\quad & \pi \, \leq \, x \, < \, 2\pi~.
                 \end{array} \right.
\end{equation}

In the limit $\Delta Y \rightarrow \infty$,
\begin{equation}
S_2 ~=~ \Sigma_2 \, + \, S_{2, \rm r}~, \quad
S_2^{\phi} ~=~ \Sigma_2^{\phi} \, + \, S_{2, \rm r}^{\phi}~.
\end{equation}
Our ratio is then
\begin{equation}
R ~=~ \frac {1} {\pi^2} \int_{\phi=0}^{2\pi} \!\!\!\!\!\! d\phi \,
g \left( \left| \phi^{\rm lab} (\phi) \, - \, \phi^{\rm lab} (\phi+\pi)
\right| \right)~, \label{eRb2}
\end{equation}
Note that this result is independent of the size of the background, $N$!
In fact, if we allow the
presence of more than one resonance, we find that $S_{2, \rm r}$ and
$S_{2, \rm r}^{\phi}$ are both proportional to the number of sources,
$n_s$, so $R$ is also independent of the number of resonances.

For the case of non--relativistic resonance decay, eq.~(\ref{eplab})
reduces to \begin{equation}
\phi^{\rm lab} = \tan^{-1} \left( \frac {(v_s / v_{\pi}) +
\sin \phi} {\cos \phi} \right) ~+~
\pi H \left( -\cos \phi \right) ~+~ \phi_r,
\label{eplnr} \end{equation}
as we can take $\cos \theta=0$.
It is clear from eqs.~(\ref{eRb2}) and (\ref{eplnr}) that, in the
non--relativistic limit, $R$ depends only on the ratio $v_s \, / \,
v_{\pi}$.  In Fig.~1, we show $R$ for $v_{\pi}=c/2$ and $v_{\pi}=c$,
along with the value in the non--relativistic limit.

In both the relativistic and non--relativistic cases, $R(0)=2$; this
is an identity, due to the different topologies of $\phi$ and $y$
($\phi=2\pi$ is equivalent to $\phi=0$, while $y=\delta y$ is not
equivalent to $y=0$).  The behavior as $v_{\pi} \rightarrow 1$ is
analytically intractable, but from Fig.~1 it is clearly not too
different from the non--relativistic case.  The limit $v_s \rightarrow
1$ can be taken analytically, yielding
\begin{equation}
R ~=~ \frac {2 \, (1-v_s^2)^{1/2}} {\pi^2} \, \ln \left[
\frac {v_s \, + \, v_{\pi}} {v_s \, - \, v_{\pi}} \right]~.
\label{eRvs1} \end{equation}
The limit of large $v_s$ differs only by a factor $\gamma_s =
(1-v_s^2)^{1/2}$ in the relativistic and non--relativistic cases.

\section{ Thermal sources } \label{sts}

A typical example of a source that emits a large number of particles is
a thermal source, such as a large droplet.  Imagine that we have a
single source, as before, but that the source emits $n_{\pi}$ particles,
where $n_{\pi} \gg 1$.  The two--particle density from this source is
\begin{equation}
\rho^{(2)}_{\rm r} (\theta_1, \phi_1, \theta_2, \phi_2) ~=~ \frac
{n_{\pi} \, (n_{\pi}-1) \, \sin \theta_1 \, \sin \theta_2} {16 \,
\pi^2}.
\end{equation}
The $\delta$--function constraints are absent, as for large $n_{\pi}$ we
can essentially ignore conservation of energy and momentum.

Again moving to the resonance rest frame, and taking $\Delta Y
\rightarrow \infty$, we obtain
\begin{equation}
S_{2, \rm r} (0) ~\propto~ n_{\pi} \, (n_{\pi}-1) \,
\int_{-\infty}^{\infty} \!\!\! dy \int_0^{2\pi} \!\!\! d\phi_1
\int_0^{2\pi} \!\!\! d\phi_2 \, \left| \frac {\partial \theta_1}
{\partial y} \right| \, \left| \frac {\partial \theta_2} {\partial y}
\right| \, \sin \theta_1 \, \sin \theta_2 ~, \label{eS2c2}
\end{equation}
and
\begin{equation}
S_{2, \rm r}^{\phi} (0) \propto 4 n_{\pi} (n_{\pi}-1)
\int_{-\infty}^{\infty} \!\!\! dy \int_{\phi_1^{\rm lab}=0}^{\pi}\!\!\!
d\phi_1 \int_{\phi_2^{\rm lab}=\pi}^{2\pi} \!\!\! d\phi_2 \left| \frac
{\partial \theta_1} {\partial y} \right| \left| \frac {\partial
\theta_2} {\partial y} \right| \, \sin \theta_1 \, \sin \theta_2.
\label{eS2pc2} \end{equation}
Taking the ratio as before, we find
\begin{equation}
R ~=~ \frac {\displaystyle 2 \int_{-\infty}^{\infty} \!\!\! dy
\int_0^{2\pi} \!\!\! d\phi_1 \int_0^{2\pi} \!\!\! d\phi_2 \, \left|
\frac {\partial \theta_1} {\partial y} \right| \, \left| \frac {\partial
\theta_2} {\partial y} \right| \, \sin \theta_1 \, \sin \theta_2 \,
g \left( \left| \phi_1^{\rm lab} \, - \, \phi_2^{\rm lab} \right|
\right)} {\pi \displaystyle \int_{-\infty}^{\infty} \!\!\! dy \left[
\int_0^{2\pi} \!\!\! d\phi \, \left| \frac {\partial \theta} {\partial
y} \right| \, \sin \theta \right]^2}~. \label{eRc2}
\end{equation}
Note that $R$ is independent of $n_{\pi}$ as well as being independent
of $N$!
In fact, $R$ is even independent of the number
of sources, as before.

We cannot evaluate the relativistic denominator for the thermal source,
but we can take the non--relativistic limit as before.  We obtain
\begin{equation}
R ~=~ \frac {1} {4 \, \pi^3} \int_0^{\pi} \!\!\! d\theta \, \sin \theta
\int_0^{2\pi} \!\!\! d\phi_1 \int_0^{2\pi} \!\!\! d\phi_2 \, g \left(
\left| \phi_1^{\rm lab} \, - \, \phi_2^{\rm lab} \right| \right)~,
\label{eRcnr} \end{equation}
where
\begin{equation}
\phi^{\rm lab} = \tan^{-1} \left( \frac {(v_s / v_{\pi}) +
\sin \theta \sin \phi} {\sin \theta \cos \phi} \right) +
\pi H \left( - \cos \phi \right) +\phi_r.
\label{eplnr2} \end{equation}
It is again clear that, in the non--relativistic limit, $R$ depends only on
$v_s \, / \, v_{\pi}$.

In Fig.~2, we display $R$ as in the case of two--particle decay.  We
find now that $R(0)=1$ and $R(\infty)=0$, so the normalization is clearly
different from that obtained for two--particle decay.  Because the
curves are different, there will be some dependence of the measured
velocity on the source size.  If the sources are known to emit many
pions, as is the case in ultra--relativistic nuclear collisions, this
dependence is weak.  However, any velocity measurement using the
proposed technique depends somewhat on knowledge of the source size.

\section{ Conclusions } \label{sc}

We have demonstrated a new procedure for measuring the ratio of a source
transverse velocity to the velocity of emitted particles, using the
split--bin correlation functions $S_2$ and $S_2^{\phi}$.  We showed the
technique for two simple cases: two--body decay, and unconstrained
decay.  In both cases, the technique gives a measurement that is
independent of the number of sources, and the background size.
In addition, we find that
the measurement is independent of the source size for large sources.
We have considered somewhat simple cases
for pedagogical purposes, in order to illustrate our technique in a
simple, model--independent framework.
The specific issue of contamination by dynamical effects should be
addressed separately.

Similar techniques could be used to measure many other quantities, so
the potential usefulness of this line of enquiry is enormous.  For
example, comparison of energy and particle--number correlation functions
could provide a relatively simple way to measure the energy distribution
of minijets in ultra--relativistic nuclear collisions.  These techniques
are probably of the most use in high multiplicity collisions, where
individual sources often cannot be resolved.  We hope that our results
will encourage other researchers to investigate new uses for correlation
functions.

\section*{ Acknowledgements }

D.S. thanks the Physics Department of McGill University for their
hospitality and computer time.  This work was partially supported by the
US Department of Energy under Grant No.\ DOE/DE-FG02-86ER-40251, by the
NSERC of Canada, and by the FCAR fund of the Quebec government.  Some
computing facilities were provided by the Ohio Supercomputer Center.

\vfill \eject

\section*{ References }

\begin{enumerate}

\item S. Pratt, Phys.\ Rev.\ D {\bf 33}, 1314 (1986); W.G. Gong {\it et
al.}, Phys.\ Rev.\ Lett.\ {\bf 65}, 2114 (1990). \label{rhbt}

\item D. Seibert, Phys.\ Rev.\ D {\bf 41}, 3381 (1990). \label{rsizes}

\item A. Capella, K. Fia\l kowski and A. Krzywicki, Phys.\ Lett.\ B
{\bf 230}, 149 (1989). \label{ranom}

\item D. Seibert, Phys.\ Rev.\ Lett.\ {\bf 63}, 136 (1989). \label{rdrops}

\item S. Voloshin and D. Seibert, Phys.\ Lett.\ B {\bf 249}, 321 (1990);
D. Seibert and S. Voloshin, Phys.\ Rev.\ D {\bf 43}, 119 (1991).
\label{rsfms}

\end{enumerate}

\vfill \eject

\section*{ Figure captions }

\begin{enumerate}

\item $R$ vs.\ $\tanh \left( v_s \gamma_s/v_{\pi} \right)$ for
two--particle decay.

\item $R$ vs.\ $\tanh \left( v_s \gamma_s/v_{\pi} \right)$ for a thermal
source.

\end{enumerate}

\vfill \eject

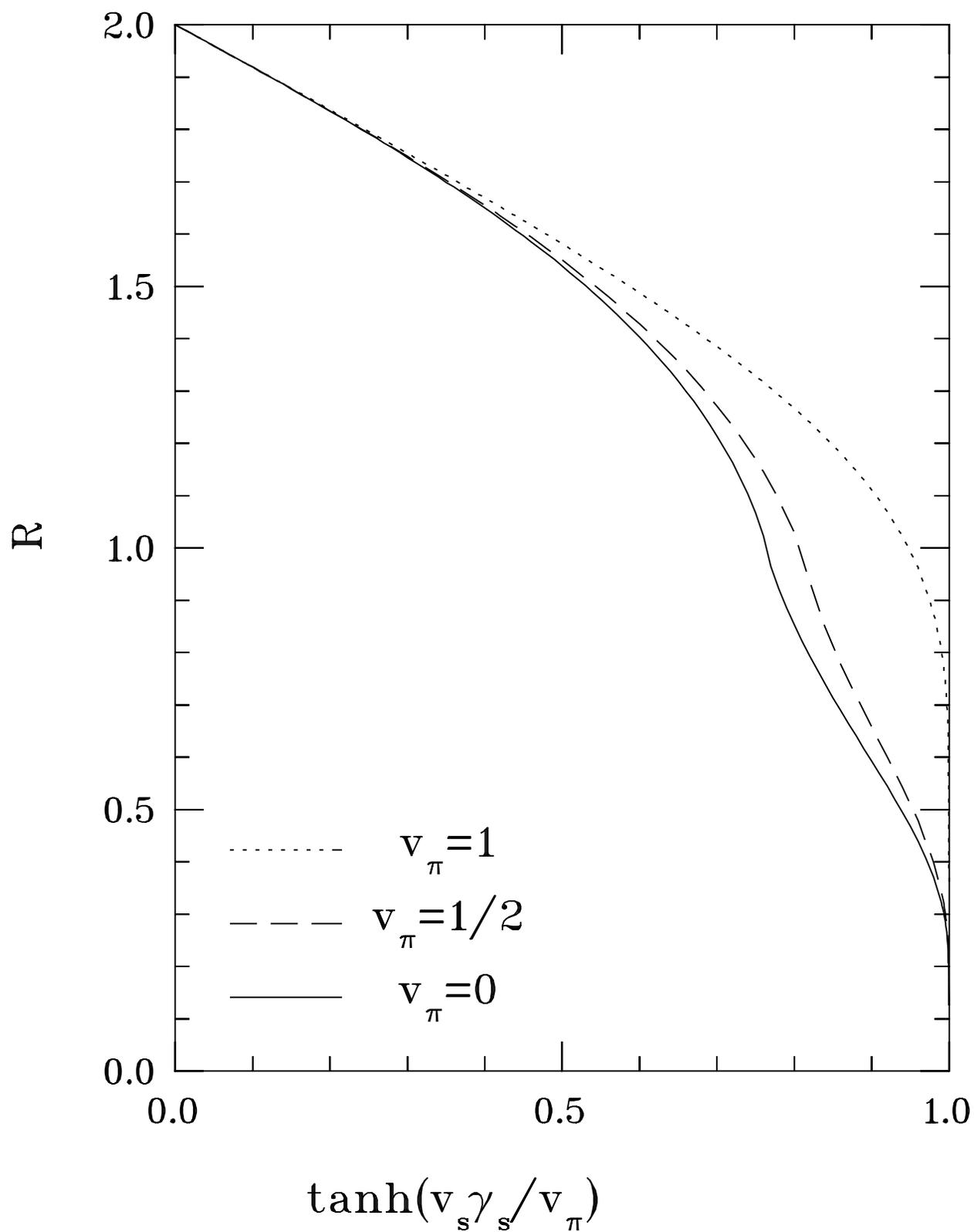
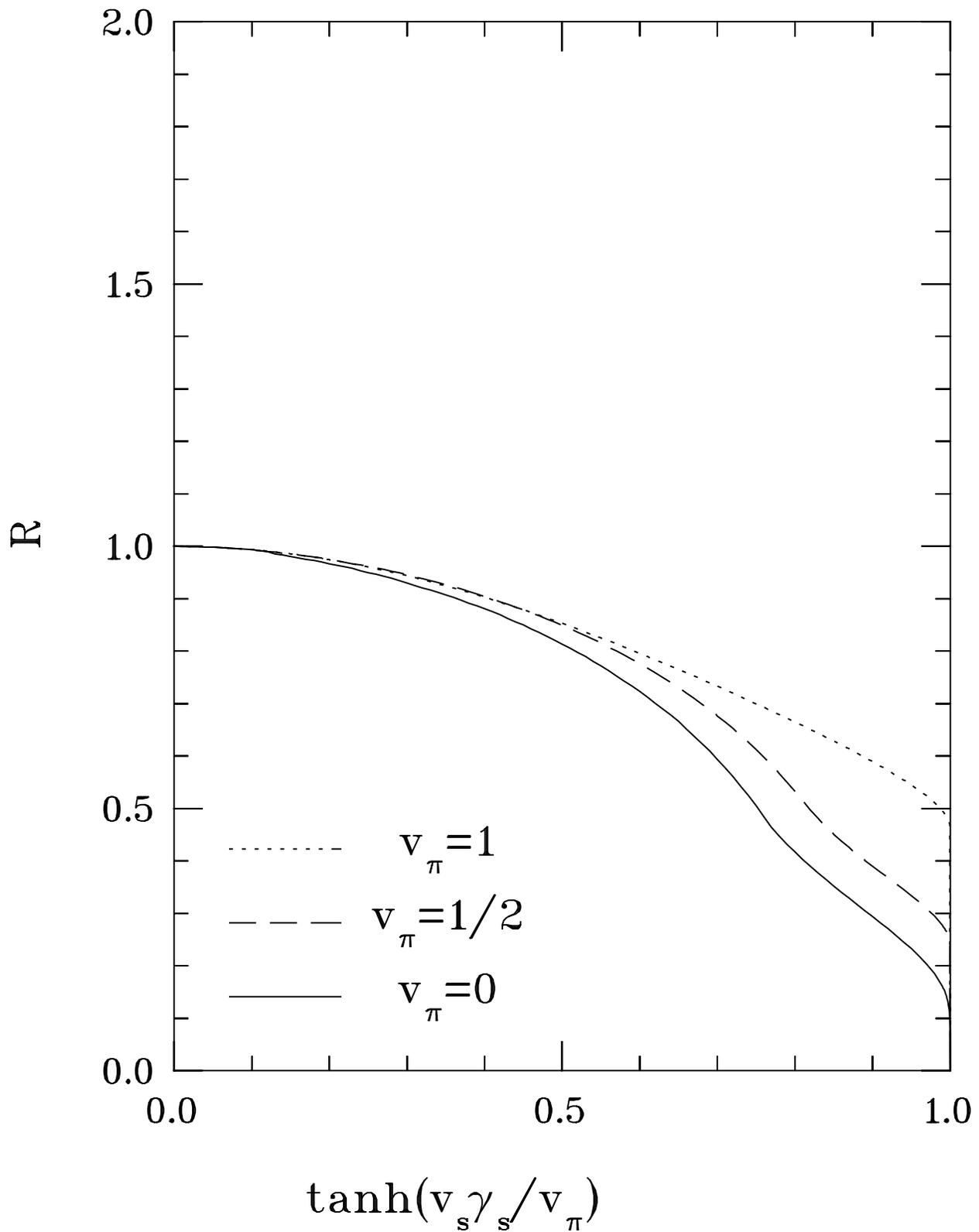
\end{document}